\documentclass[sigconf]{acmart}
\usepackage{multirow}
\renewcommand\footnotetextcopyrightpermission[1]{}
\AtBeginDocument{%
  }

\begin{document}

\title{UITrans: Seamless UI Translation from Android to HarmonyOS}

\author{Lina Gong}
\affiliation{%
  \institution{Nanjing University of Aeronautics and Astronautics}
  \city{Nanjing}
  \country{China}}
\email{gonglina@nuaa.edu.cn}
\orcid{0000-0002-5272-6706}
\author{Chen Wang}
\affiliation{%
  \institution{Nanjing University of Aeronautics and Astronautics}
  \city{Nanjing}
  \country{China}}
\email{wangchen712@nuaa.edu.cn}

\author{Yujun Huang}
\affiliation{%
  \institution{Nanjing University of Aeronautics and Astronautics}
  \city{Nanjing}
  \country{China}}
\email{yujunhuang@nuaa.edu.cn}

\author{Di Cui}
\affiliation{%
  \institution{Xidian University}
  \city{Xi'an}
  \country{China}
}
\email{cuidi@xidian.edu.cn}

\author{Mingqiang Wei}
\affiliation{%
  \institution{Nanjing University of Aeronautics and Astronautics}
  \city{Nanjing}
  \country{China}}
\email{mqwei@nuaa.edu.cn}

\renewcommand{\shortauthors}{Lina et al.}

\begin{abstract}
  Seamless user interface (i.e., UI) translation has emerged as a pivotal technique for modern mobile developers, addressing the challenge of developing separate UI applications for Android and HarmonyOS platforms due to fundamental differences in layout structures and development paradigms. In this paper, we present UITrans, the first automated UI translation tool designed for Android to HarmonyOS. UITrans leverages an LLM-driven multi-agent reflective collaboration framework to convert Android XML layouts into HarmonyOS ArkUI layouts. It not only maps component-level and page-level elements to ArkUI equivalents but also handles project-level challenges, including complex layouts and interaction logic. Our evaluation of six Android applications demonstrates that our UITrans achieves translation success rates of over 90.1\%, 89.3\%, and 89.2\% at the component, page, and project levels, respectively. UITrans is available at https://github.com/OpenSELab/UITrans and the demo video can be viewed at https://www.youtube.com/watch?v=iqKOSm\\CnJG0.

\end{abstract}

%\begin{CCSXML}
%<ccs2012>
 % <concept_id>00000000.0000000.0000000</concept_id>
 % <concept_desc>Do Not Use This Code, Generate the Correct Terms for Your Paper</concept_desc>
%  <concept_id>00000000.00000000.00000000</concept_id>
  %<concept_desc>Do Not Use This Code, Generate the Correct Terms for Your Paper</concept_desc>
%  <concept_significance>300</concept_significance>
 % <concept_id>00000000.00000000.00000000</concept_id>
  %<concept_desc>Do Not Use This Code, Generate the Correct Terms for Your Paper</concept_desc>
%  <concept_significance>100</concept_significance>
 %</concept>
 % <concept_id>00000000.00000000.00000000</concept_id>
%  <concept_desc>Do Not Use This Code, Generate the Correct Terms for Your Paper</concept_desc>
 % <concept_significance>100</concept_significance>
%</ccs2012>
%\end{CCSXML}

\ccsdesc[500]{Software and its engineering}
\ccsdesc[300]{Software and its engineering~Software Translation}
%\ccsdesc{Do Not Use This Code~Generate the Correct Terms for Your Paper}
%\ccsdesc[100]{Do Not Use This Code~Generate the Correct Terms for Your Paper}

\keywords{UI Translation, Android, HarmonyOS, LLMs}
 % \Description{Enjoying the baseball game from the third-base
 % \label{fig:teaser}


\maketitle

\section{Introduction}
As mobile development platforms continue to diversify, developers face significant challenges in deploying applications across multiple platforms. Traditionally, this requires developing separate applications for each platform, resulting in higher development costs and longer iteration cycles~\cite{biorn2018survey, karami2023impact}. A more efficient approach is to utilize translation techniques that allow applications to be transferred from one platform to another, minimizing the need for platform-specific development efforts. However, user interfaces (UIs) translation across platforms presents substantial challenges, primarily due to fundamental differences in layout structures and development paradigms~\cite{feng2022autoicon, morgado2015impact}.

\begin{figure}[htbp]
\centerline{\includegraphics[width=0.9\linewidth, scale=0.9]{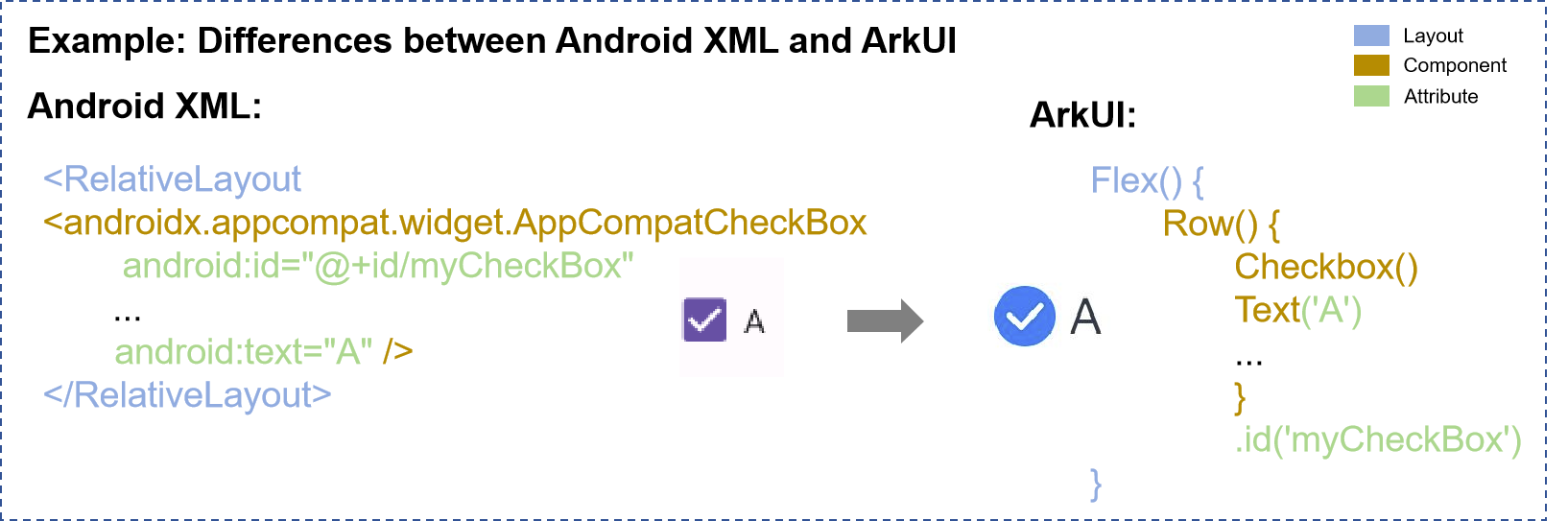}}
\caption{An example of differences between Android XML and ArkUI.}
\label{fig:Difference}
\end{figure}

This issue is particularly evident when comparing Android and HarmonyOS, two leading mobile operating systems with fundamentally different UI frameworks. Android uses an XML-based layout system, while HarmonyOS employs the ArkUI framework. Figure ~\ref{fig:Difference} illustrates a UI example on both Android and HarmonyOS, highlighting differences in layout structure, components, and attributes. As a result, Android applications cannot be run directly on HarmonyOS devices, necessitating UI translation for compatibility~\cite{li2023openharmony, wang2011research, ostrander2012android, marchenko2023jetpack}. 

Manual UI translation from Android to HarmonyOS is often inefficient and error-prone, especially for complex mobile applications. This is further exacerbated by the relatively limited number of HarmonyOS-compliant developers.  Consequently, there is a clear and pressing need for an automated UI translation solution. While some existing tools attempt semi-automated UI translation with rule-based libraries from Android to IOS platform~\cite{gao2024uimigration}, they are typically limited to simpler UI elements and do not address the complexities of large layouts, interaction logic, or third-party library compatibility between Android and HarmonyOS. %Furthermore, these tools often fail to ensure the completeness and consistency of converted applications at the project level.

To address these challenges, we propose UITrans, the first automated UI translation tool designed for Android to HarmonyOS. UITrans leverages an LLM-driven multi-agent reflective
collaboration framework to convert Android XML layouts into HarmonyOS ArkUI layouts. Specifically, UITrans utilizes LLM to analyze the structure and semantics of Android XML layouts, while employing the Retrieval-Augmented Generation (i.e., RAG) combined with decision rules to map components, attributes, and events to their ArkUI counterparts. More importantly, UITrans achieves precise translation of complex layouts and interaction logic through the integration of a reflective optimization mechanism~\cite{renze2024selfreflection}. 

We conducted experiments on six open source Android
projects and 32 UI pages from GitHub to evluate the effectiveness of UITrans. Experimental results demonstrate
that UITrans performs exceptionally well in component, page, and project-level UI translation, significantly reducing the
developers’ workload. Additionally, we are the first to provide a benchmarks for the UI translation from Android to HarmonyOS, including component, page, and project-level dataset, which would serve as evaluation benchmark for future practitioners and researchers. 

%The primary contributions of this paper are as follows:

%A2H Converter leverages a large language model (LLM)-based multi-agent framework, enhanced with reflective collaboration techniques. The tool utilizes LLM to analyze the structure and semantics of Android XML layouts and employs the RAG combined with decision rules library to map components, attributes, and events to their ArkUI counterparts\cite{5}. The multi-agent collaboration framework, augmented with reflective optimization, ensures the accurate conversion of complex layouts and interaction logic\cite{7}.

%These architectural differences create significant barriers to seamless UI translation. In particular, the static hierarchical XML layouts on Android contrast with ArkUI’s dynamic code-driven component model. As a result, Android applications cannot be run directly on HarmonyOS devices, requiring developers to convert Android UIs for compatibility with HarmonyOS\cite{1}. 

\section{Our Approach}

\begin{figure*}[htbp]
\centerline{\includegraphics[width=0.9\linewidth, scale=0.9]{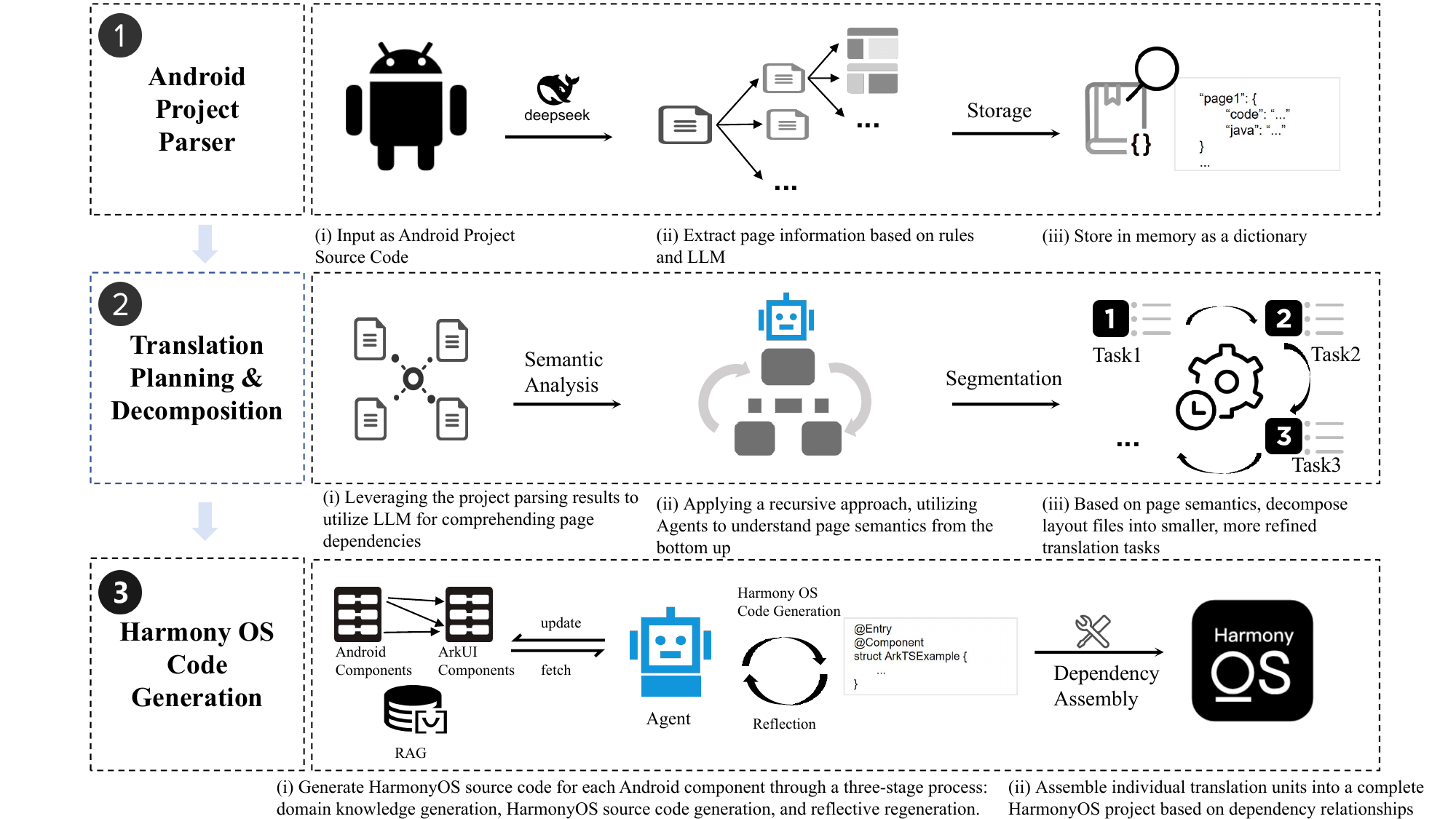}}
\caption{The architecture of UITrans.}
\label{fig:overview}
\end{figure*}

UITrans's architecture is depicted in Figure~\ref{fig:overview}, which takes the source code of Android project as input and the source code of HarmonyOS as output. The core of UITrans relies on a multi-agent reflective collaboration mechanism~\cite{hong2023metagpt, shinn2024reflexion, singh2024agentic, park2023generative}, powered by a large language model, to parse Android components, convert them into intermediate functional representations, and ultimately generate HarmonyOS code.

UITrans is divided into three main components: Android Project Parsing, Translation Task Planning and Decomposition, and HarmonyOS Code Generation. Specifically,

\begin{itemize}
\item {\texttt{Android Project Parsing}}: is responsible for analyzing the logical structure of the application, including both pages and subpages, to prepare the application for the migration process (see Section~\ref{sec:parse} for details).
\item {\texttt{Translation Task Planning and Decomposition}}: employs systematic decomposition rules to identify and isolate the smallest translatable units, allowing for an organized and efficient translation strategy (see Section~\ref{sec:task_plan}).
\item {\texttt{HarmonyOS Code Generation}}: utilizes a multi-agent collaboration mechanism, enhanced by the RAG and UI mapping table, to translate Android components into ArkUI components. Then it reassembles the generated ArkUI components in accordance with the logical structure provided by the parsing module, thus generating the complete HarmonyOS source code (see Section~\ref{sec:harmonyOS}).
\end{itemize}

Next, we would describe all these components hereafter.

\subsection{Android Project Parsing}\label{sec:parse}

During the Android Project Parsing phase, UITrans performs a page-level analysis, where each page represents a distinct user interface visible at a given time~\cite{liu2018uiscript,nam2024codeunderstanding}. UITrans conducts an in-depth examination of the complex interaction logic within the project, including interactions between pages and between main pages and subpages. Meanwhile, for further processing, we store these interaction logic in a dictionary format.

Specifically, UITrans first extracts all activity-related information from the application's manifest file (\textit{AndroidManifest.xml}), where each activity corresponds to an independent user interface. The tool then locates the Java code and XML layout files associated with each activity. Following this, the tool parses the Java and XML code by analyzing component references and XML syntax (e.g., the \textit{findViewById()} method in Java files and \textit{<include>} tags in XML files). This process allows the identification of XML dependencies, components, and associated resource files.

Meanwhile, to address potential gaps in interaction logic arising from incomplete rule-based parsing\cite{li2024staticanalysis}, UITrans leverages the DeepSeek (V2.5)~\footnote{\url{https://huggingface.co/deepseek-ai/DeepSeek-V2.5}} to supplement the parsing results~\cite{liu2024deepseek}. This enhances the tool's ability to comprehensively extract all main pages, subpages, and their corresponding code snippets for each activity. Additionally, we provide the prompt used for Android project parsing with DeepSeek in our GitHub~\footnote{\url{https://github.com/OpenSELab/UITrans/blob/master/images/prompts/parser_prompt.png}}, including XML layout analysis, Java File analysis and unified requirements.

\begin{figure}[htbp]
\centerline{\includegraphics[width=0.9\linewidth, scale=0.9]{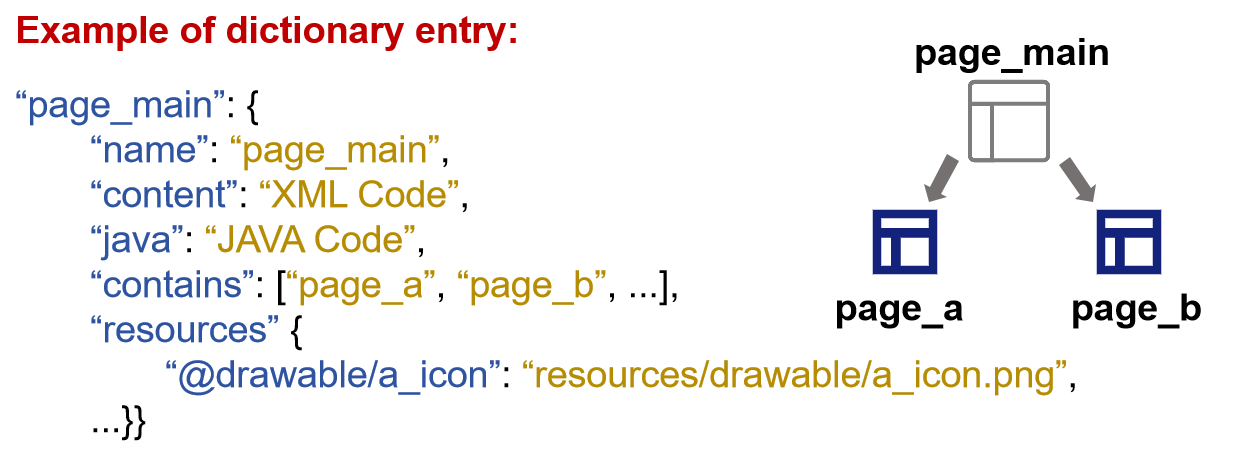}}
\caption{An example of dictionary entry.}
\label{fig:Dict}
\end{figure}

Based on these parsing and dependency analyses, an object is constructed to encapsulate the Java code, XML layout code, page dependencies, and resource file indices—providing a comprehensive representation of each page. These objects are stored in memory in a dictionary format (presented in Figure~\ref{fig:Dict}), thereby enabling efficient use in the subsequent translation and generation phases.

\subsection{Translation Task Planning \& Decomposition}\label{sec:task_plan}

During the Translation Task Planning \& Decomposition phase, UITrans employs a recursive approach to decompose complex translation tasks into smaller, manageable units, thereby enhancing both translation accuracy and efficiency. Specifically, the process begins by analyzing the page dependency relationships identified during the parsing phase, enabling the identification of interaction logic between main pages and subpages. Subsequently, the tool leverages multiple LLM-driven agents to semantically interpret the pages in a bottom-up manner, breaking down layout files into smaller translatable units, thereby providing a clear roadmap for HarmonyOS code generation. Meanwhile, we provide the prompt used for  translation task planning \& decomposition with DeepSeek in our GitHub~\footnote{\url{https://github.com/OpenSELab/UITrans/blob/master/images/prompts/breakdown_layout.png}}, including Android
XML layout and breakdown rules. %The prompt used for breakdown layout with DeepSeek is shown in Figure~\ref{fig:combined} (a) of Appendix~\ref{sec:translation_prompt}.

To ensure effective processing of these smaller translation units by the LLM-driven agents, UITrans defines a set of decomposition rules. Furthermore, to address the challenges posed by differences in component mapping, attribute and style definitions, and custom widget compatibility between Android XML layouts and HarmonyOS ArkUI, UITrans uses LLM-driven agents to generate functional descriptions for each translatable unit (the prompts used for description generation with DeepSeek are provided in our GitHub~\footnote{\url{https://github.com/OpenSELab/UITrans/blob/master/images/prompts/generate_component_description.png}}). These descriptions are created using natural language processing techniques, enabling accurate matching and generation, thereby ensuring that the final translated output faithfully reflects the original application's functionality.

%Figure~\ref{fig:combined} (b) of Appendix~\ref{sec:translation_prompt} shows the used prompt

\subsection{HarmonyOS Code Generation}\label{sec:harmonyOS}

This phase consists of two major steps: the generation of minimal translation units and the assembly of their dependencies. These steps are performed concurrently through the use of a multi-agent reflective collaboration mechanism, enabling a comprehensive project-level translation from Android XML layouts to ArkUI components. To mitigate inconsistencies introduced by platform differences and enhance both the accuracy and efficiency of the translation, UITrans utilizes two core resources: the UI Mapping Translation Table and the RAG Knowledge~\cite{bhattarai2024fewshot, mansourian2024iosandroid}.

The UI Mapping Translation Table defines the mapping relationships between Android XML components and their ArkUI counterparts. This table (provided in GitHub repository~\footnote{\url{https://github.com/OpenSELab/UITrans/blob/master/script/initialize_database/component_table.json}})) includes detailed component descriptions, code of source Android component, code of target component, usage examples, and functional specifications. The RAG Knowledge (provided in GitHub repository~\footnote{\url{https://github.com/OpenSELab/UITrans/blob/master/script/initialize_database/components.json}})), on the other hand, offers comprehensive documentation for ArkUI components, encompassing functional descriptions, attributes, and examples, all of which facilitate the translation process by ensuring correct understanding and generation of corresponding ArkUI code. Additionally, we use the bge-m3 model~\footnote{\url{https://huggingface.co/BAAI/bge-m3}} to generate vector embeddings for the documents in RAG, which are stored in the Chroma vector database~\footnote{\url{https://www.trychroma.com/}}.

\subsubsection{Generation of Minimal Translation Units}

The generation of minimal translation units involves converting individual Android components into their corresponding HarmonyOS source code. This process is carried out in three distinct stages:

\textbf{Step 1. Domain Knowledge Generation:} The primary objective of domain knowledge generation is to provide precise and accurate support for generating HarmonyOS source code. UITrans first queries the UI Mapping Translation Table to identify potential mappings between Android components and their ArkUI counterparts. If a matching component is found, the system uses the BM25 combined with Ranker algorithm to rank the results and selects the top three most relevant candidates. Among this, the BM25 algorithm is used for sparse matching and Reranker is employed to reorder the results, which reduces noise and ensures that the generated code accurately reflects the original functionality. The mapping results, including usage examples and source code of Android component, serve as critical domain knowledge, aiding the LLM in generating HarmonyOS-compatible code~\cite{li2023retrieval}. Additionally, we provide the prompt used for the query of domain knowledge in our GitHub~\footnote{\url{https://github.com/OpenSELab/UITrans/blob/master/images/prompts/generate_component_query.png}}, including ules and query instructions to ensure accurate component mapping and
retrieval of relevant documentation or examples.

In cases where no direct mapping exists in the table, UITrans utilizes the model’s inference capabilities. Leveraging deep semantic understanding, the model can infer potential HarmonyOS substitutes and retrieve relevant attributes and usage examples from the RAG Knowledge, ensuring that appropriate source code of HarmonyOS component is generated even when no direct match is available.
%AdditionFigure~\ref{fig:combined} (C) of Appendix~\ref{sec:translation_prompt} presents the prompt used in this process.

\textbf{Step 2. HarmonyOS Source Code Generation:} After domain knowledge is generated, UITrans enters the code generation phase. The system uses the domain knowledge along with the Android source code as input to guide LLM in generating corresponding ArkUI code for HarmonyOS. The model not only generates the appropriate ArkUI components but also adjusts the code based on the functional and structural requirements of the Android components, ensuring that the code is fully compatible with HarmonyOS while maintaining a consistent user experience in comparison to the original Android application. Additionally, we provide the prompt used for  HarmonyOS code generation in our GitHub~\footnote{\url{https://github.com/OpenSELab/UITrans/blob/master/images/prompts/translation_prompt.png}}.

\textbf{Step 3. Reflective Re-generation:} To enhance the accuracy of the generated code, UITrans introduces a reflective re-generation process. A secondary LLM-driven agent compares the generated code with the original functional description, identifying any inconsistencies or deviations. These discrepancies are then fed back to the LLM, which uses this feedback to optimize the code in the next iteration of generation. This reflective, iterative process ensures that the generated code maintains functional consistency, structural integrity, and overall accuracy. Furthermore, for components where no direct mapping could be identified, the system records these mappings in the UI Mapping Translation Table, enriching the table’s content for future use. Additionally, we provide the prompt used for triggering reflective re-generation of LLM-driven agent in our GitHub~\footnote{\url{https://github.com/OpenSELab/UITrans/blob/master/images/prompts/reflection_prompt.png}}.

\subsubsection{Dependency Assembly}

After the source code for all minimal translation units is generated, UITrans proceeds to the project assembly phase. During this stage, UITrans utilizes the page dependencies and interaction logic extracted during the project parsing phase to reorder the generated ArkUI component code according to the correct sequence and logical structure, ensuring the construction of a complete HarmonyOS application project. The system guarantees the accuracy of page interactions, component dependencies, and resource references during the assembly, ensuring that the translated application maintains the same layout as the original Android application and functions seamlessly on the HarmonyOS platform. Meanwhile, we provide the prompt for assemble components in our GitHub~\footnote{\url{https://github.com/OpenSELab/UITrans/blob/master/images/prompts/assemble_components.png}}.

\section{Tool Implementation}

We have released a publicly accessible online service~\footnote{\url{http://124.70.54.129:37860/}}, accompanied by a video demonstration). The technology stack includes Gradio (frontend), Python (backend translation program and web services), SQLite (for storing translation tables), and ChromaDB (for storing and retrieving corpora). This systematic approach enables the efficient and accurate use of our UITrans tool to perform UI translation from Android to HarmonyOS, covering component, page, and project level of Android projects. Users simply upload the Android code to our service,  the corresponding HarmonyOS code is generated through clicking "Confirm" button.

\section{Preliminary empirical Evaluation}

In this section, we present preliminary empirical results that assess the effectiveness of UITrans on UI translation. We selected the top six Android applications that could be successfully built, compiled, and previewed the UI from GitHub across different domains based on their star count. As showed in Table~\ref{tab:projects} of Appendix~\ref{appendix_data}, these applications span diverse domains, including education, science, and e-commerce, with varying page counts, component types, and total lines of code (ranging from 164 to 1861 lines). Meanwhile, we define complex components as those other than native Android components and the commonly used components added to the UI mapping Translation table.%This selection ensures a fair and representative evaluation of the tool's performance across different use cases. For each application, we successfully built, compiled, and previewed the UI to validate the migration results.

% \begin{table*}[h]
% \centering
% \caption{Android project dataset}\label{tab:projects}
% \small
% \begin{tabular}{lccccccc} 
% \hline
% \textbf{Project} & \textbf{Domain} & \textbf{Pages} & \textbf{Components} & \textbf{Complex\%} & \textbf{Comp. Success\%} & \textbf{Page Success\%} & \textbf{Project Success\%} \\
% \hline
% Bookdash Android & Education & 5 & 31/16 & 25.0\% & 94.1\%(22) & 93.8\%(3) & 92.0\% \\
% Transportr & Travel & 2 & 15/9 & 0.0\% & 90.1\%(10) & 90.5\%(2) & 89.2\% \\
% PSLab Android App & Science & 4 & 41/11 & 25.0\% & 97.3\%(36) & 94.9\%(4) & 96.5\%  \\
% enjoyshop & E-commerce & 4 & 46/10 & 50.0\% & 91.1\%(43) & 91.8\%(3) & 90.7\% \\
% GoGrocery & E-commerce & 13 & 164/17 & 30.7\% & 95.2\%(103) & 94.4\%(5) & 93.7\% \\
% forecastie & Daily Life & 4 & 44/6 & 25.0\% & 94.6\%(33) & 89.3\%(0) & 89.2\% \\
% \hline
% \end{tabular}
% \begin{flushleft}
% \textit{Note:} The values in parentheses indicate the count of components and pages that were successfully migrated without any modifications.
% \end{flushleft}
% \end{table*}

%: \textbf{component migration success rate (i.e., (Comp. Success))}, \textbf{page migration success rate (i.e., Page Success)}, and \textbf{project migration success rate (i.e., Project Success)} as our evaluated metrics. These metrics are 

Furthermore, we define three key metrics \textbf{Comp. Success}, \textbf{Page Success}, and \textbf{Project Success} as our evaluated metrics. These metrics are defined in the Appendix~\ref{appendix_metric}. Meanwhile, we compare our UITrans with a direct prompt-based LLM approach in Table~\ref{tab:comparison}.

%as follows:

% \textbf{Component Migration Success Rate} = Average(1 -proportion of code lines requiring modification per component)

% \textbf{Page Migration Success Rate} = Average(1 -proportion of code lines requiring modification per page)

% \textbf{Project Migration Success Rate} = 1 - (lines of code requiring modification at the project level) / (total lines of code in the translated project)

\begin{table}[h]
\centering
\caption{Experimental results between the onlyPrompt and A2H methods}\label{tab:comparison}
\resizebox{\linewidth}{!}{
\begin{tabular}{l|l ccc}
\toprule
\textbf{Project}          & \textbf{Method}  & \textbf{Comp. Success\%} & \textbf{Page Success\%} & \textbf{Project Success\%} \\ 
\midrule
\multirow{2}{*}{\textbf{Bookdash Android}} 
    & Only-prompt   & 42.2\%(3)                 & 50.6\%(0)                  & 43.8\%                    \\ 
    & UITrans           & \textbf{94.1\%(22)}         & \textbf{93.8\%(3)}         & \textbf{92.0\%}           \\ 
\cmidrule(l){1-5}
\multirow{2}{*}{\textbf{Transportr}}        
    & Only-prompt   & 35.4\%(0)                  & 40.1\%(0)                  & 42.2\%                    \\ 
    & UITrans            & \textbf{90.1\%(10)}         & \textbf{90.5\%(2)}         & \textbf{89.2\%}           \\ 
\cmidrule(l){1-5}
\multirow{2}{*}{\textbf{PSLab Android App}} 
    & Only-prompt   & 52.0\%(5)                  & 59.5\%(1)                  & 53.3\%                    \\ 
    & UITrans           & \textbf{97.3\%(36)}         & \textbf{94.9\%(4)}         & \textbf{96.5\%}           \\ 
\cmidrule(l){1-5}
\multirow{2}{*}{\textbf{enjoyshop}}         
    & Only-prompt   & 38.2\%(3)                  & 42.9\%(0)                  & 41.8\%                    \\ 
    & UITrans           & \textbf{91.1\%(43)}         & \textbf{91.8\%(3)}         & \textbf{90.7\%}           \\ 
\cmidrule(l){1-5}
\multirow{2}{*}{\textbf{GoGrocery}}         
    & Only-prompt   & 37.3\%(9)                  & 41.8\%(1)                  & 38.9\%                    \\ 
    & UITrans            & \textbf{95.2\%(103)}         & \textbf{94.4\%(5)}         & \textbf{93.7\%}           \\ 
\cmidrule(l){1-5}
\multirow{2}{*}{\textbf{forecastie}}        
    & Only-prompt   & 37.0\%(5)                  & 41.4\%(0)                  & 41.3\%                    \\ 
    & UITrans            & \textbf{94.6\%(33)}         & \textbf{89.3\%(0)}         & \textbf{89.2\%}           \\ 
\bottomrule
\end{tabular}
}
\begin{flushleft}
\textit{Note:} The values in parentheses indicate the count of components and pages that were successfully built and compiled without any modifications.
\end{flushleft}
\end{table}

From Table~\ref{tab:comparison}, we observe that compared with only prompt method, UITrans achieves a success rate of over 89\% at the component, page, and project levels, with the PSLab Android App project reaching as high as 96.5\%. Additionally, more than 72.5\% of components were successfully translated without requiring any modifications. More importantly, the modified translated code~\footnote{\url{https://github.com/OpenSELab/UITrans/tree/master/examples/modified_examples}} primarily involves types of attribute usage or alignment or positioning of layout in the specific component, which can also be easily fixed by newcomers.%In contrast, the success rate of migration using a direct prompt-based large model approach reached a maximum of XXX across the component, page, and project levels. \textbf{These results demonstrate that A2H Converter effectively and accurately performs UI migration from Android to HarmonyOS, significantly reducing manual effort and associated costs.}

\section{Conclusion}

In this paper, we present UITrans, the
first automated UI translation tool designed for Android to HarmonyOS at the component, page and project levels. UITrans has been applied and validated in different Android applications, which achieves a success rate of over 89.3\%. Meanwhile, we have released a publicly accessible online service with simplified operations.

%A2H Converter leverages the semantic understanding and inference capabilities of large language models, alongside a multi-agent reflective feedback mechanism, to achieve a fully automated and efficient translation process. By utilizing these advanced AI-driven techniques, the tool is capable of handling complex, project-level layout files, providing a comprehensive solution for UI migration. Experiments on six Android applications collected from GitHub demonstrate that our A2H Converter achieves a migration success rate of over 89.3\% at the component, page, and project levels. A2H Converter could automatically analyze, map, and generate equivalent ArkUI components for HarmonyOS ensures both accuracy and consistency, thereby allowing developers to focus on more value-added tasks rather than repetitive, error-prone manual adjustments.
\begin{acks}
This study is supported by the National Natural Science Foundation of China under Grant No. 62202223, the Natural Science Foundation of Jiangsu Province, China under Grant No. BK20220881, and the Fundamental Research Funds for the Central Universities, NO. NJ2024029.
\end{acks}

\bibliographystyle{ACM-Reference-Format}
% \bibliographystyle{unsrt}
\bibliography{software}

\appendix

\section{The Benchmark of Android projects}\label{appendix_data}

We present the top six Android applications that could be successfully built, compiled, and previewed the UI as follows:

\begin{table}[h]
\centering
\caption{Android project dataset}\label{tab:projects}
\resizebox{\linewidth}{!}{%
\begin{tabular}{lcccc} 
\hline
\textbf{Project} & \textbf{Domain} & \textbf{Pages} & \textbf{Components} & \textbf{Complex\%} \\ 
\hline
Bookdash Android & Education & 5 & 31/16 & 25.0\% \\
Transportr & Travel & 2 & 15/9 & 0.0\% \\
PSLab Android App & Science & 4 & 41/11 & 25.0\% \\
enjoyshop & E-commerce & 4 & 46/10 & 50.0\% \\
GoGrocery& E-commerce & 13 & 164/17 & 30.7\% \\
forecastie & Daily Life & 4 & 44/6 & 25.0\% \\
\hline
\end{tabular}
}
\begin{flushleft}
\renewcommand{\arraystretch}{1.0}  
\small
\textit{Note:} Each value in \textbf{Components} column respectively represent the total number of components (before the slash) and the number of complex components (after the slash) within the pages.
\end{flushleft}
\end{table}

\section{The definition of evaluation metrics}\label{appendix_metric}

To evaluate the effectiveness of our UITrans, we propose component migration success rate, page migration success rate, and project migration success rate metrics. These metrics are defined as follows:

%\noindent\textbf{Component Migration Success Rate :}
\[
\scriptsize
\text{Comp. Success} = 
\frac{1}{n} \sum_{i=1}^{n} 
\left( 1 - \frac{\text{Lines Modified}_i}{\text{Total Lines}_i} \right)
\]

%\noindent\textbf{Page Migration Success Rate :}
\[
\scriptsize
\text{Page Success} = 
\frac{1}{m} \sum_{j=1}^{m} 
\left( 1 - \frac{\text{Lines Modified}_j}{\text{Total Lines}_j} \right)
\]

%\noindent\textbf{Project Migration Success Rate ():}
\[
\scriptsize
\text{Project Success} = 
1 - \frac{\text{Lines Modified at Project Level}}{\text{Total Lines in Project}}
\]

Among them, the variables used in the formulas are defined as follows:

\begin{itemize}
    \item \(n\): Number of components in the project.
    \item \(m\): Number of pages in the project.
    \item \(\text{Lines Modified}_i\): Lines of code requiring modification in the \(i\)-th component.
    \item \(\text{Total Lines}_i\): Total lines of code in the \(i\)-th component.
    \item \(\text{Lines Modified}_j\): Lines of code requiring modification in the \(j\)-th page.
    \item \(\text{Total Lines}_j\): Total lines of code in the \(j\)-th page.
    \item \(\text{Lines Modified at Project Level}\): Total lines of code requiring modification for the entire project.
    \item \(\text{Total Lines in Project}\): Total lines of code in the translated project.
\end{itemize}

Note that the Lines Modified represent the number of modified code lines ensuring that the translated HarmonyOS code is functionally consistent with the Android UI. 

\end{document}